\documentclass[twocolumn,showpacs,preprintnumbers,amsmath,amssymb]{revtex4}
\usepackage{dcolumn}
\usepackage{bm}
\usepackage[dvips]{graphicx}
\usepackage{wrapfig,subfigure}
\usepackage{amssymb}
\usepackage{color}

\newcommand{\n}{\mbox{\boldmath $\nabla$}}
\newcommand{\qb}{{\bf q}}
\newcommand{\pb}{{\bf p}}
\newcommand{\fb}{{\bf f}}
\newcommand{\cb}{{\bm \chi}}

\newcommand{\qbo}{{\bf q}_{\rm opt}}
\newcommand{\qbon}{{\bf q}_{n\,\rm opt}}

\newcommand{\dotqbon}{\dot{\bf q}_{n\,\rm opt}}

\newcommand{\pbon}{{\bf p}_{n\,\rm opt}}

\begin{document}

\title{Resonant symmetry lifting in a parametrically modulated oscillator}
\author{ D. Ryvkine and M. I. Dykman}
\affiliation{
Department of Physics and Astronomy, Michigan State
 University, East Lansing, MI 48824, USA}
\date{\today}

\begin{abstract}
We study a parametrically modulated oscillator that has two stable
states of vibrations at half the modulation frequency $\omega_F$.
Fluctuations of the oscillator lead to interstate switching.  A
comparatively weak additional field can strongly affect the
switching rates, because it changes the switching activation
energies. The change is linear in the field amplitude. When the
additional field frequency $\omega_d$ is $\omega_F/2$, the field
makes the populations of the vibrational states different thus
lifting the states symmetry. If $\omega_d$ differs from
$\omega_F/2$, the field modulates the state populations at the
difference frequency, leading to fluctuation-mediated wave mixing.
For an underdamped oscillator, the change of the activation energy
displays characteristic resonant peaks as a function of frequency.

\end{abstract}


\maketitle

\section{Introduction}
\label{sec:Introduction}

A parametrically modulated oscillator is one of the simplest
physical systems that display spontaneous breaking of
time-translation symmetry. When the modulation is sufficiently
strong, the oscillator has two states of vibrations at half the
modulation frequency, the period two states \cite{LL_Mechanics2004}.
They are identical except for the phase shift by $\pi$, but for each
of them the symmetry with respect to time translation by the
modulation period is broken. Fluctuations of the oscillator lead to
switchings between the states. Switching rates are equal by
symmetry, for stationary fluctuations. The switchings ultimately
make the state populations equal, thus restoring the full
time-translation symmetry. Experimental studies of
fluctuation-induced switchings in classical parametrically modulated
systems were done for trapped electrons \cite{Lapidus1999},
optically trapped atoms \cite{Kim2005,Kim2006}, and
microelectromechanical systems \cite{Chan_parametric}. The obtained
switching rates are in good agreement with the theory
\cite{Dykman1998}.

The degeneracy of period two states can be lifted if, in addition to
parametric modulation at frequency $\omega_F$, a system is driven at
frequency $\omega_F/2$. In the frame oscillating at frequency
$\omega_F/2$ the period two states and the additional field look
static. The system reminds an Ising ferromagnet, with the period two
states and the additional field playing the roles of spin
orientations and an external magnetic field, respectively. One can
expect that the role of the direction of the magnetic field is
played by the phase of the additional field counted off from the
phase of one of the period two states. Depending on the field phase,
one or the other state should be predominantly occupied.

In this paper we study resonant symmetry lifting in a parametric
oscillator, which occurs where the frequency of the additional field
is close to the oscillator eigenfrequency. The field makes the rates
of switching between the period two states, $W_{12}$ and $W_{21}$,
different from each other. In turn, this leads to the difference of
the stationary state populations. This difference may become large
even for a comparatively weak field, as can be surmised from the
analogy with the problem of a ferromagnet. There the change of the
state populations becomes large once the energy difference of the
states due to an external magnetic field exceeds $kT$, which happens
already for weak fields where this difference itself is small
compared to the internal energy.

In contrast to the case of a ferromagnet, the energy of a
parametrically modulated oscillator is not conserved and its
stationary distribution is not of the Boltzmann form. However if the
fluctuation intensity is small, the dependence of the switching
rates on this intensity is often of the activation type, see
Refs.~\onlinecite{Freidlin_book},~\onlinecite{Dykman2001} and papers
cited therein. This applies not only to a classical, but also to a
quantum oscillator, where switching is due to quantum fluctuations
\cite{Marthaler2006}. An additional field changes the effective
switching activation energies, and once this change exceeds the
fluctuation intensity, the overall change of the switching rates
becomes large. The effect is particularly strong if the field is
resonant.

In what follows we develop a theory of the switching rates $W_{nm}$
for a classical oscillator. We find the dependence of the switching
activation energies on the amplitude, phase, and frequency of the
additional field and on the oscillator parameters. We first study
the symmetry lifting by the field at frequency
$\omega_d=\omega_F/2$. Of particular interest here is the vicinity
of the bifurcation point where the period two states merge, as in
this range the rates $W_{nm}$ are comparatively large and easy to
control.

We are also interested in the situation where the additional field
frequency $\omega_d$ is not exactly equal to $\omega_F/2$. Here, the
field-induced modulation of the switching probabilities causes
oscillations of the state populations at frequency
$|\omega_d-\omega_F/2|$. Such oscillations, superimposed on the
oscillator vibrations at frequency $\omega_F/2$ in period two
states, lead to vibrations at frequency $|\omega_F-\omega_d|$, i.e.,
to a strong effective fluctuation-induced three-wave mixing.

The amplitude of the population oscillations becomes small when
$|\omega_d-\omega_F/2|$ largely exceeds the switching rates. Still
the rates themselves may be significantly changed by the additional
field. Of primary interest in this case are the rates $W_{nm}$
averaged over the period $4\pi/|2\omega_d-\omega_F|$ and their
dependence on the additional field amplitude $A_d$ and frequency
$\omega_d$. One may expect that, as in the case of equilibrium
systems \cite{Larkin1986,Ivlev1986,Linkwitz1991}, the rate change is
quadratic in $A_d$ for small $A_d$ and corresponds to effective
heating of the system by the field.  For underdamped equilibrium
systems this heating can be resonant, as seen for modulated
Josephson junctions \cite{Devoret1987,Turlot1998}. However, for
somewhat stronger fields the change of the logarithm of the
switching rate should become linear in $A_d$
\cite{Dykman1997,Smelyanskiy1997c}. This happens when the properly
scaled field amplitude exceeds temperature.

We show that, for a parametric oscillator, the activation energies
are indeed linear in the additional field amplitude $A_d$ once it is
not too small (but is also not too large). When the oscillator is
underdamped, the factor multiplying $A_d$ displays a characteristic
resonant frequency dependence. We develop a technique that allows us
to find this dependence in an explicit form, for the considered
nonequilibrium system. The asymptotic analytical results are
compared with the results of numerical calculations of the
activation energies.

In Sec.~II we discuss the Langevin equation for a parametrically
modulated nonlinear oscillator in the rotating frame and give the
general expression for the probability of switching between
coexisting stable vibrational states. In Sec.~III we obtain a
general expression for the correction to the activation energy of
switching and show that it is linear in $A_d$. We study resonant
symmetry lifting of the switching rates. In Sec.~IV we investigate
low-frequency oscillations of the state populations and
fluctuations-mediated resonant wave mixing. In Sec.~V an expression
for the period-averaged switching rate is given. In Sec.~VI we
consider symmetry lifting and the frequency dependence of the
activation energy close to the bifurcation point where the period
two states merge. In Sec.~VII we study the case of weak damping. We
show that the change of the activation energy may display
characteristic asymmetric resonant peaks as a function of the
additional field frequency. Sec.~VIII contains concluding remarks.

\section{Langevin equation and the switching rates}
\label{sec:Dynamics}

We will study switching between period two states of a nonlinear
oscillator, which is parametrically modulated by a force
$F\cos(\omega_F t)$ and is additionally driven by a comparatively
weak field $A_d\cos(\omega_d t+\phi_d)$ at frequency
$\omega_d\approx \omega_F/2$. The Hamiltonian of the oscillator is a
sum of the term that describes the motion without the extra field
and the term proportional to the field, $H_{\rm osc}= H_{\rm
osc}^{(0)} + H_{\rm osc}^{(d)}$,
\begin{eqnarray}
\label{eq:H_0(t)} H_{\rm
osc}^{(0)}&=&\frac{1}{2}p_0^2+\frac{1}{2}q_0^2\left(\omega_0^2+F\cos(\omega_F
t)\right)+\frac{1}{4}\gamma q_0^4,\nonumber\\
H_{\rm osc}^{(d)}&=& -q_0A_d\cos(\omega_d t+\phi_d)
\end{eqnarray}
($q_0$ and $p_0$ are the coordinate and momentum of the oscillator).
We will assume that the modulation frequency $\omega_F$ is close to
twice the frequency of small amplitude vibrations $\omega_0$, and
that the driving force $F$ is not too large so that the oscillator
nonlinearity remains small,
\begin{eqnarray}
\label{eq:resonant_conditions}
\left|\omega_F-2\omega_0\right|,\,|\omega_d-\omega_0| \ll\omega_0\, ,\\
\quad F\ll\omega_0^2\, ,\quad |\gamma|\langle q^2 \rangle
\ll\omega_0^2.\nonumber
\end{eqnarray}
In what follows for concreteness we set $\gamma>0$.

Following the standard procedure \cite{LL_Mechanics2004}, we change
to the rotating frame and introduce the dimensionless canonical coordinate Q
and momentum P,
\begin{eqnarray}
\label{eq:P_Q_varibales}
 q_0(t)&=&C\left[P\cos(\omega_F t/2)-Q\sin(\omega_F t/2)\right],\\
 p_0(t)&=&-C\frac{\omega_F}{2}\left[P\sin(\omega_F t/2)+Q\cos(\omega_F
 t/2)\right] ,\nonumber
\end{eqnarray}
where $C=(2F/3\gamma)^{1/2}$. In these variables the Hamiltonian
becomes equal to $\tilde H_{\rm osc}=(F^2/6\gamma)g(Q,P)$, with
$g=g^{(0)}+g^{(d)}(\tau)$,
\begin{eqnarray}\label{eq:g_function}
g^{(0)}&=&\frac{1}{4}\left(P^2+Q^2\right)^2+\frac{1}{2}(1-\mu)P^2-
\frac{1}{2}(1+\mu)Q^2,\nonumber\\
g^{(d)}(\tau)&=&-a_d\left[P\cos\left(\nu_d\tau +
\phi_d\right)\right.\nonumber\\
&&\left.+Q\sin\left(\nu_d\tau + \phi_d\right)\right].
\end{eqnarray}
Here we introduced dimensionless time $\tau$ and dimensionless
parameters $\mu$ and $\nu_d$. These parameters characterize,
respectively, the detuning of the modulation frequency from the
oscillator eigenfrequency and the detuning of the weak-field
frequency from $\omega_F/2$, i.e., an effective ``beat frequency"
with the subharmonic of the strong field,
\begin{eqnarray}
\label{eq:dimensionless} \mu=\omega_F(\omega_F-2\omega_0)/F,&&\quad
\nu_d
=\omega_F(2\omega_d-\omega_F)/F,\nonumber\\
\tau&=&tF/2\omega_F.
\end{eqnarray}
The parameter $a_d=A_d(6\gamma/F^3)^{1/2}$ is the dimensionless
amplitude of the additional driving field. In obtaining
Eq.~(\ref{eq:g_function}) we used the rotating wave approximation
and disregarded fast oscillating terms $\propto \exp(\pm i n\omega_F
t)\, ,\,n\geq 1 $. In the quantum formulation, the eigenvalues of
$g^{(0)}(Q,P)$ give the scaled quasienergy of the system
\cite{Marthaler2006}, and in what follows for brevity we call $g$
quasienergy.

We will assume that the interaction with a bath that leads to
dissipation of the oscillator is sufficiently weak, so that the
oscillator is underdamped. Then under fairly general assumptions
\cite{Dykman1979a}, in the rotating frame  dissipation is described
by an instantaneous friction force (no retardation). Also the noise
spectrum is generally practically flat in a comparatively narrow
frequency range of width $\gtrsim F/\omega_F$ centered at
$\omega_0$; this is the most interesting range, since the oscillator
filters out noise at frequencies far from this range. Therefore with
respect to the slow time $\tau$ the noise can be assumed white. The
oscillator motion is described by the Langevin equation, which can
be conveniently written in a vector form as
\begin{eqnarray}
\label{eq:Langevin}
 \dot {\bf q}\equiv \frac{d{\bf q}}{d\tau}={\bf K}+
 {\bf f}(\tau), \qquad {\bf K}={\bf K}^{(0)}+{\bf K}^{(d)},
\end{eqnarray}
with
\begin{eqnarray}
\label{eq:K_force}
 {\bf K}^{(0)}=-\zeta^{-1}{\bf q} +\hat\epsilon\n
 g^{(0)},\qquad
 {\bf K}^{(d)}(\tau)=\hat\epsilon\n
 g^{(d)}(\tau).
\end{eqnarray}
Here all vectors have two components, ${\bf q}\equiv (Q,P)$, ${\bf
K}\equiv (K_Q,K_P)$, and $\n\equiv (\partial_Q,\partial_P)$, while
$\hat\epsilon$ is the permutation tensor,
$\epsilon_{QQ}=\epsilon_{PP}=0$, $\epsilon_{QP}=-\epsilon_{PQ}=1$.
The parameter $\zeta^{-1}$ in Eq.~(\ref{eq:Langevin}) gives the
oscillator friction coefficient in the units of $F/2\omega_F$. We
use for $\zeta$ and $\mu$ the same notations as in
Refs.~\onlinecite{Dykman1998,Marthaler2006}. Because of the extra
field $\propto a_d$, the function ${\bf K}^{(d)}$ explicitly depends
on time.

The function ${\bf f}(\tau)$ is a random force. Its two components
are independent white Gaussian noises with the same intensity
\cite{Dykman1979a},
\[\langle f_Q(\tau)f_Q(0)\rangle = \langle f_P(\tau)f_P(0)\rangle =
2D\delta(\tau-\tau').\]
The noise intensity $D$ is the smallest parameter of the theory. If
both the friction force and the noise come from coupling to a
thermal reservoir at temperature $T$, we have $D=6\gamma
kT/F\zeta\omega_F^2$ (the parameter $D$ corresponds to
$D\zeta^{-2}/2$ in Ref.~\onlinecite{Dykman1998}). .

\subsection{Oscillator dynamics in the absence of noise}
\label{subsection:noise_free_dynamics}

In the absence of noise and the symmetry-breaking field $\propto
A_d$, in the range
\begin{equation}
\label{eq:mu_B} \mu_B^{(1)} < \mu < \mu_B^{(2)}, \qquad
\mu_B^{(1,2)} = \mp(1-\zeta^{-2})^{1/2}
\end{equation}
the parametrically modulated oscillator has two stable period-two
states ${\bf q}_{1,2}^{(0)}$ and an unstable state ${\bf
q}_b^{(0)}$. These states are the stationary solutions of equation
${\bf K}^{(0)}=0$. They merge for $\mu=\mu_B^{(1)}$. The stable
states 1, 2 are inversely symmetrical, ${\bf q}_2^{(0)}=-{\bf
q}_1^{(0)}$. For concreteness we choose
\[Q_1^{(0)}=-Q_2^{(0)} >0.\]
The vibration amplitude in the unstable state is zero,
$\qb_b^{(0)}={\bf 0}$. For $\mu>\mu_B^{(2)}$ the state ${\bf q=0}$
becomes stable and there additionally emerge two unstable period two
states \cite{LL_Mechanics2004}.

We will be interested in a comparatively weak symmetry-breaking
field. Respectively, we will assume that the reduced field amplitude
$a_d$ is small, so that the field does not lead to new stable
states. It just makes the stationary states periodic, for $\nu_d\neq
0$, or shifts them, for $\nu_d=0$. The correspondingly modified
states are given by the periodic solutions of equation $\dot\qb={\bf
K}$ or by equation ${\bf K}={\bf 0}$. In the laboratory frame the
periodic stable states ${\bf q}_{1,2}(\tau)$ correspond to
oscillator vibrations at frequency $\omega_F/2$ weakly modulated at
frequency $|\omega_F-2\omega_d|/2$. They have spectral components at
$\omega_F/2,\omega_d$, and the ``mirror" frequency
$|\omega_F-\omega_d|$.

We will limit ourselves to the analysis of switching in the
parameter range (\ref{eq:mu_B}). In this range escape from a period
two state leads to switching to a different period two state. For
$\mu > \mu_B^{(2)}$ escape may result in a transition to the
zero-amplitude state (from which the system may also escape to one
of the period two states). The results of the paper immediately
extend to this range, but this extension will not be discussed.
Therefore we use the terms ``escape" and ``switching"
intermittently.

\subsection{Switching rates: General formulation}
\label{section:switching_probabilities}

The noise $\fb(t)$ leads to fluctuations about the stable states and
to interstate transitions. When the noise is weak, fluctuations
have small amplitude on average. Interstate transitions require
large outbursts of noise and therefore occur infrequently. For
Gaussian $\delta$-correlated noise the probability of a transition
from $n$th to $m$th stable period two state has activation
dependence on the noise intensity $D$ and is given by the expression
\cite{Freidlin_book,Dykman2001}
\begin{eqnarray}
\label{eq:switching_general} &&W_{nm}=C_W\exp(-R_n/D),\\
R_n&=&\min\int\nolimits_{-\infty}^{\infty} d\tau L\left(\dot{\bf
q},{\bf q};\tau\right), \qquad L=\frac{1}{4}(\dot{\bf q}-{\bf
K})^2.\nonumber
\end{eqnarray}
The quantity $R_n$ is the activation energy of a transition. It is
given by the solution of a variational problem. The minimum in
Eq.~(\ref{eq:switching_general}) for $R_n$ is taken with respect to
trajectories ${\bf q}(\tau)$ that start for $\tau\to -\infty$ at the
initially occupied stable state ${\bf q}_n(\tau)$ and asymptotically
approach ${\bf q}_b(\tau)$ for $\tau\to\infty$. The optimal
trajectory $\qbon(t)$ that minimizes $R_n$ is called the most
probable escape path (MPEP). The system is most likely to follow
this trajectory in a large fluctuation that leads to escape.

The prefactor in the transition probability is $C_W\propto
F/\omega_F$ in unscaled time $t$. It weakly depends on the noise
intensity and will not be discussed in what follows.

The variational problem (\ref{eq:switching_general}) for $R_n$ can
be associated with the problem of dynamics of an auxiliary system
with Lagrangian $L$ and coordinates ${\bf q}=(Q,P)$, and with
Hamiltonian
\begin{eqnarray}
\label{eq:Hamiltonian}
 &&H=H^{(0)} + H^{(d)},\qquad H^{(0)}={\bf p}^2 +
{\bf pK}^{(0)}, \\
&&H^{(d)}=\pb{\bf K}^{(d)}.\nonumber
\end{eqnarray}
In terms of this auxiliary Hamiltonian system, the MPEP of the
original dissipative system corresponds to a heteroclinic trajectory
that goes from the periodic (or stationary, for $\nu_d=0$) state
$({\bf q}_n(\tau), {\bf p=0})$ to the periodic state $({\bf
q}_b(\tau),{\bf p=0})$.

\section{Perturbation theory for the activation energy}
\label{sec:perturbed_activation_energy}

When the additional field $\propto a_d$ is weak, the activation
energy $R_n$ is close to its value $R^{(0)}$ for $a_d=0$, which is
the same for the both states $n=1,2$. Even though the correction to
$R^{(0)}$ is small compared to $R^{(0)}$, it may significantly
exceed the noise intensity $D$. Then the overall change of the
transition probability $W_{nm}$ will be exponentially large.

The correction to the activation energy was studied earlier for
thermal equilibrium systems additionally modulated by a
comparatively weak periodic field
\cite{Dykman1997,Smelyanskiy1997c}. As mentioned in the
Introduction, it was found that the correction to $R_n$ is linear in
the field amplitude. The factor multiplying the amplitude gives the
logarithm of the transition probability and therefore was called the
logarithmic susceptibility (LS). We show now that the correction to
the activation energy for a parametrically modulated oscillator is
also linear in the amplitude of the additional field, $\delta
R_n=R_n-R^{(0)}\propto a_d$, and find the proportionality
coefficient, that is the LS.

Because of the modulation at frequency $\omega_F$, the oscillator is
far away from thermal equilibrium even without the field $\propto
a_d$. This leads to a significant difference of the problem of
switching from that for equilibrium systems. Its physical origin is
the lack of time reversibility in nonequilibrium systems. As a
result the auxiliary Hamiltonian system described by the Hamiltonian
$H^{(0)}$ (\ref{eq:Hamiltonian}), which gives optimal fluctuational
trajectories and the MPEP of the original dissipative system, is
nonintegrable \cite{Graham1984b}.

The Hamiltonian dynamics described by the full Hamiltonian $H$
(\ref{eq:Hamiltonian}), which includes the additional time-dependent
field $\propto a_d$, is also nonintegrable. However, for small $a_d$
the heteroclinic Hamiltonian trajectory $\qbon(\tau)$ that gives the
MPEP may remain close to the unperturbed heteroclinic trajectory
$\qbon^{(0)}(\tau)$ for $a_d=0$. Then, as in the case of modulated
equilibrium systems \cite{Dykman1997,Smelyanskiy1997c}, the
lowest-order correction in $a_d$ to the variational functional $R_n$
(\ref{eq:switching_general}) can be found by calculating the
perturbing term in the Lagrangian along $\qbon^{(0)}(\tau)$.

The trajectory $\qbon^{(0)}(\tau)$ is a real-time counterpart of an
instanton \cite{Langer1967} and is often called a real-time
instanton. It goes from $\tau\to -\infty$ to $\tau\to\infty$. As in
the case of standard instantons, $\dotqbon^{(0)}(\tau)$ looks like a
pulse. It is large only for a time of the order of the relaxation
time of the system in the absence of noise. The position of the
center of the pulse on the time axis $\tau_c$ is arbitrary.

Periodic modulation lifts the degeneracy with respect to $\tau_c$.
It synchronizes switching events. It is this synchronization that
leads to the correction to $R_n$ being linear in $a_d$ for a nonzero
frequency detuning $|\nu_d|$. The synchronization corresponds to
calculating the field-induced correction $\delta R_n$ using as a
zeroth order approximation the unperturbed MPEP for the $n$th state
$\qbon^{(0)}(\tau-\tau_c)$ and adjusting $\tau_c$ in such a way that
the overall functional $R_n=R^{(0)} +\delta R_n$ be minimal. That
is, $\tau_c$ is found from the condition of the minimum of the
function $\delta R_n(\tau_c)$,
\begin{eqnarray}
\label{eq:deltaR_n}
 \delta R_n&=&\min_{\tau_c}\delta R_n(\tau_c),\\
 \delta R_n(\tau_c)&=&-\int\nolimits_{-\infty}^{\infty}d\tau{\bm
 \chi}_n(\tau-\tau_c){\bf K}^{(d)}(\tau),\nonumber\\
 {\bm\chi}_n(\tau)&=&\frac{1}{2}\left[\dotqbon^{(0)}(\tau) -
 {\bf K}^{(0)}\left(\qbon^{(0)}(\tau)\right)\right].\nonumber
\end{eqnarray}
Clearly, $\tau_c$ is determined modulo the dimensionless modulation
period $2\pi/|\nu_d|$. Generically, there is one MPEP per period
that provides the absolute minimum to $R_n$.

Equation (\ref{eq:deltaR_n}) shows that the correction to the
activation energy of escape is indeed linear in $a_d$. The
coefficient at $a_d$ is determined by the Fourier components
$\tilde\cb_n(\pm\nu_d)$ of the function ${\bm \chi}(\tau)$,
\begin{eqnarray}
\label{eq:chi_omega}
 \tilde{\cb}_n(\nu)=\int\nolimits_{-\infty}^{\infty}d\tau\,
 e^{i\nu\tau}{\bm \chi}_n(\tau).
\end{eqnarray}
The function $\cb_n(\tau)$ and its Fourier transform
$\tilde\cb_n(\nu)$ give the LS in the time- and frequency
representation. They are of central interest for the studies of
symmetry lifting and switching rate modulation. Interestingly, the
functions $\cb_n(\tau), \tilde\cb_n(\nu)$ have two components each,
even though there is only one additional force that drives the
oscillator. This leads to important consequences discussed below and
is in contrast to the case of a modulated equilibrium system.

We note that the condition $d[\delta R_n(\tau_c)]/d\tau_c=0$,
Eq.~(\ref{eq:deltaR_n}), corresponds to the condition that the
Mel'nikov function ${\cal M}(\tau_c)$ defined for our Hamiltonian
system with two degrees of freedom as
\begin{eqnarray}
\label{eq:Melnikov_function}
 {\cal M}(\tau_c)=\int\nolimits_{-\infty}^{\infty}d\tau
 \left\{H^{(0)}\left(\tau-\tau_c\right),\,
 H^{(d)}(\tau-\tau_c;\tau)\right\},
\end{eqnarray}
be equal to zero. Here, $\{A,B\}$ is the Poisson bracket with
respect to the dynamical variables of the auxiliary system ${\bf
q,p}$. The Poisson bracket is evaluated along the unperturbed
trajectory $\qbon^{(0)}(\tau-\tau_c),\pbon^{(0)}(\tau-\tau_c)$.
Respectively, the functions $\qb,\pb$ in $H^{(0)}, H^{(d)}$ are
evaluated at time $\tau-\tau_c$, which is indicated by the first
argument in $H^{(0)}, H^{(d)}$; the second argument in $H^{(d)}$
indicates the explicit time dependence of $H^{(d)}$,
Eq.~(\ref{eq:Hamiltonian}), due to the field $\propto a_d$. For
systems with one degree of freedom, the condition ${\cal
M}(\tau_c)=0$ shows that $\tau_c$ for the unperturbed trajectory is
chosen in such a way that this trajectory is close to the trajectory
$\qbon(\tau-\tau_c), \pbon(\tau-\tau_c)$ \cite{Guckenheimer1987}.
For our system the condition ${\cal M}(\tau_c)=0$ is necessary for
the applicability of perturbation theory. The corresponding analysis
will be provided elsewhere.

\subsection{Symmetry lifting by a field at subharmonic frequency}
\label{subsec:symmetry_lifting}

The analysis of the switching rates should be done somewhat
differently in the case where the frequencies of the additional
field and the parametrically modulating field satisfy the condition
$\omega_d=\omega_F/2$. In this case $\nu_d=0$ and the force ${\bf
K}^{(d)}$ is independent of time. Therefore the function $\delta
R_n(\tau_c)$ is independent of $\tau_c$ and minimizing it over
$\tau_c$ is irrelevant. However, a perturbation theory in $a_d$ can
still be developed. The first order correction to the activation
energy $\delta R_n\equiv\delta R_n^{\rm res}(\phi_d)$ can be
obtained by evaluating the term $\propto a_d$ in the Lagrangian $L$
(\ref{eq:switching_general}) along the unperturbed path
$\qbon^{(0)}(\tau)$. It has a simple explicit form
\begin{eqnarray}
\label{eq:staticR_n}
 \delta R_n^{\rm res}(\phi_d)=a_d\left[\tilde\chi_{nQ}(0)\cos\phi_d
 -\tilde\chi_{nP}(0)\sin\phi_d\right],
\end{eqnarray}
where the subscripts $Q,P$ enumerate the components of $\tilde\cb$
[we have also used Eqs.~(\ref{eq:g_function}), (\ref{eq:K_force})
for ${\bf K}^{(d)}$]. Both components of $\tilde\cb_n(0)$ contribute
to the change of the activation energy for subharmonic driving.

An important feature to emphasize is that the function $\delta
R_n^{\rm res}(\phi_d)$ is {\em not} the limit of $\min\delta
R_n(\tau_c)$ for $\nu_d \to 0$. Physically this is because
Eq.~(\ref{eq:deltaR_n}) gives the change of the logarithm of the
escape rate $W_{nm}$ averaged over the period
$\pi/|2\omega_d-\omega_F|$ (or $2\pi/|\nu_d|$, in dimensionless time
$\tau$). Such averaging is meaningful as long as the frequency
detuning $|2\omega_d-\omega_F|\gg W_{nm}$. In the opposite limit the
occupation of the state $n$ changes significantly over the period
$\pi/|2\omega_d-\omega_F|$; this change is not characterized by the
period-averaged $W_{nm}$.

An important property of the static LS $\tilde\cb_n(0)$ is that it
has opposite signs for the states 1 and 2. Indeed, these states are
inversely symmetric, ${\bf q}_1^{(0)}=-{\bf q}_2^{(0)}$, and since
the vector ${\bf K}^{(0)}$ is also antisymmetric, ${\bf K}^{(0)}\to
-{\bf K}^{(0)}$ for ${\bf q}\to -{\bf q}$, it is clear that
$\qb^{(0)}_{1\,\rm opt}(\tau)=- \qb^{(0)}_{2\,\rm opt}(\tau)$, and
therefore $\tilde{\bm \chi}_1(0)=-\tilde{\bm \chi}_2(0)$. Then from
Eqs.~(\ref{eq:switching_general}), (\ref{eq:staticR_n}) we have for
the switching rates $W_{12}$ and $W_{21}$
\begin{eqnarray}
\label{eq:prob_ratio_static}
 W_{12}/W^{(0)}=W^{(0)}/W_{21}=\exp\left[-\delta R_1^{\rm res}(\phi_d)/D\right],
\end{eqnarray}
where $W^{(0)}=W^{(0)}_{12}= W^{(0)}_{21}$ is the switching rate for
$a_d=0$, $W^{(0)}\propto\exp(-R^{(0)}/D)$.
Eq.~(\ref{eq:prob_ratio_static}) applies for arbitrary $a_d/D$. The
corrections to the prefactor $\propto a_d$ (but not $a_d/D$) have
been disregarded.

In the parameter range $\mu_B^{(1)} < \mu < \mu_B^{(2)}$ where the
only stable states of the system are the period two states [cf.
Eq.~(\ref{eq:mu_B})], the ratio of the stationary populations of the
states $w_1$ and $w_2$ is the inverse of the escape probabilities
ratio. Therefore from Eq.~(\ref{eq:prob_ratio_static})
\[w_1/w_2=W_{21}/W_{12}=\exp\left[2\delta R_1^{\rm res}(\phi_d)/D\right].\]

It is seen from this expression that even a comparatively weak
symmetry-lifting field, where $|\delta R_1^{\rm res}|\ll R^{(0)}$,
can lead to a significant change of the state populations. This
happens when $|\delta R_1^{\rm res}|\gg D$. The ratio of the state
populations is determined by the phase of the field $\phi_d$. As
mentioned in the Introduction, there is a similarity with magnetic
field induced symmetry breaking for an Ising spin, with $\phi_d$
playing the role of the orientation of the magnetic field. We note
that both the absolute value and the {\it sign} of the coefficients
$\tilde{\bm \chi}_n(0)$ depend also on the oscillator parameters for
$a_d=0$. Therefore one can change the population ratio not only by
varying the phase and amplitude of the additional field, but also by
varying these parameters, for example the amplitude $F$ of the
parametrically modulating field.

\section{Low-frequency modulation of state populations}

The LS $\tilde\cb_n(\nu)$ displays frequency dispersion for $\nu$ of
the order of the inverse dimensionless relaxation time of the
oscillator or higher. If the additional field is so closely tuned to
the subharmonic frequency that $|2\omega_d-\omega_F|\lesssim
W^{(0)}$, the period averaging of $W_{nm}$ implied in
Eq.~(\ref{eq:deltaR_n}) becomes inapplicable, as explained above.
One should rather think of the instantaneous values of the switching
rates $W_{nm}(\tau)$, which are given by
Eq.~(\ref{eq:prob_ratio_static}) with $\phi_d\to \nu_d\tau +
\phi_d(0)$. Slow time-dependent modulation of the switching rates
leads to modulation of the state populations,
\begin{eqnarray}
\label{eq:population_modulation}
 \frac{dw_1}{d\tau}=-W_{12}(\tau)w_1 + W_{21}(\tau)w_2,\qquad w_1+w_2=1.
\end{eqnarray}
Such modulation has attracted much attention in the context of
stochastic resonance \cite{Dykman1995d,Wiesenfeld1998}. In contrast
to a particle in a slowly modulated double-well static potential
that has been most frequently studied in stochastic resonance, here
the stable states are fast oscillating, and strong modulation of
their populations is induced by a high-frequency driving field. In
this sense there is similarity with stochastic resonance in a
resonantly driven oscillator with coexisting period one states (see
Ref.~\onlinecite{Dykman1995d}) that was recently observed in
experiment \cite{Chan2006}. In contrast to a resonantly driven
oscillator, for a parametrically modulated oscillator the
populations of period two states are equal in the absence of extra
field for any parameter values.

Oscillations of the state populations at frequency $|\nu_d|$
($|\omega_d-\omega_F/2|$ in the laboratory frame) have a
comparatively large amplitude $\propto a_d/D$, for small noise
intensity. They lead to vibrations of the oscillator in the
laboratory frame at frequencies $\omega_d$ and
$|\omega_F-\omega_d|$, with the average coordinate being
\begin{eqnarray}
\langle q_0(t)\rangle &\approx&
|2F/3\gamma|^{1/2}\left[w_1(t)-w_2(t)\right]\nonumber\\
&&\times\left[P_1^{(0)}\cos(\omega_Ft/2)-Q_1^{(0)}\sin\omega_Ft/2\right].\nonumber
\end{eqnarray}
The vibration amplitude is much larger than the amplitude of
noise-free vibrations about attractors. This indicates
strong fluctuation-induced effective three-wave mixing. Moreover,
because oscillations of $w_{1,2}(\tau)$ are nonsinusoidal for large
$a_d/D$, there also occurs multiple-wave mixing. Further analysis of
this effect is beyond the scope of the present paper.

\section{Period-averaged switching rate}
\label{subsec:average_rate}

For the difference frequency $|2\omega_d-\omega_F|\gg W_{nm}^{(0)}$
the major effect of the additional field is the change of the
switching rate averaged over the dimensionless period
$2\pi/|\nu_d|$. It is given by the change of the activation energy
$\delta R_n$. From Eq.~(\ref{eq:deltaR_n}) it follows that $\delta
R_n$ is independent of time. From the symmetry relations
$\qb^{(0)}_{1\,\rm opt}(\tau)=- \qb^{(0)}_{2\,\rm opt}(\tau)$ and
${\bf K}^{(d)}(\tau)=-{\bf K}^{(d)}\left(\tau+\pi\nu_d^{-1}\right))$
it follows that the minimization over $\tau_c$ in
Eq.~(\ref{eq:deltaR_n}) leads to $\delta R_1=\delta R_2$. It is
straightforward to show that
\begin{eqnarray}
\label{eq:high-frequency}
 &&\delta R_1=\delta R_2=-a_d\tilde\chi_{1c}(\nu_d), \\
 &&\tilde\chi_{1c}(\nu)=\left[|\tilde{\cb}_1(\nu)|^2-
 i\tilde\cb_1(\nu)\hat\epsilon\tilde\cb_1^*(\nu)\right]^{1/2}.\nonumber
 \end{eqnarray}

From Eq.~(\ref{eq:high-frequency}), the change of the activation
energy is fully determined by the LS at the scaled frequency
difference $\nu_d$. The function $\tilde\chi_{1c}$ is nonnegative.
It displays a characteristic dependence on the oscillator
parameters. As we show, it may have resonant peaks in the regime of
small damping, $\zeta \gg 1$.

\section{Scaling behavior near the bifurcation point}
\label{sec:close_to_bifurcation}

\subsection{Symmetry lifting}

The dynamics of the oscillator is simplified near the bifurcation
point $\mu_B^{(1)}$  where the period two states merge together (a
supercritical pitchfork bifurcation \cite{Guckenheimer1987}). Here,
motion is controlled by one slow variable (soft mode) $Q'=Q\cos\beta
+ P\sin\beta$, where $\beta =\frac{1}{2}(\pi -\arcsin \zeta^{-1})$.
From Eq.~(\ref{eq:Langevin}), to the lowest order in the distance to
the bifurcation point $\eta=\mu-\mu_B^{(1)}$ the Langevin equation
for this variable has the form
\begin{eqnarray}
\label{eq:slow_variable}
 &&\dot Q'=-\partial_{Q'} U + f'(\tau),
 \qquad
 U= \frac{1}{2}\mu_B^{(1)}\eta\zeta Q'^2
 \nonumber\\
 &&\quad -\frac{1}{4}\mu_B^{(1)}\zeta^3Q'^4
 +Q'a_d\cos(\nu_d\tau+\phi_d+\beta),
 \end{eqnarray}
where $f'(\tau)$ is white noise of intensity $2D$.

The potential $U$ in the absence of additional field has a familiar
form of a quartic parabola (note that $\mu_B^{(1)}<0$). The values
of the slow variable at the period two states correspond to the
minima of $U$, $Q_{1,2}^{\prime\,(0)}=\pm \eta^{1/2}\zeta^{-1}$,
whereas the unstable zero-amplitude state is at the local maximum of
$U$ at $Q^{\prime\,(0)}=0$. The activation energy
$R^{(0)}=\mu_B^{(1)}\eta^2/4\zeta$ is just the height of the
potential barrier $\Delta U$
 \cite{Dykman1998}; the probability distribution near the pitchfork
bifurcation point was discussed in
Refs.~\nocite{Knobloch1983}\nocite{Graham1987a}\cite{Knobloch1983,Graham1987a}.

For an additional field at exact subharmonic frequency, the change
of the activation energy is simply the change of the barrier height.
With the chosen convention that $Q_1^{(0)}>0$ we obtain from
Eq.~(\ref{eq:slow_variable})
\begin{equation}
\label{eq:deltaR_bif_static}
 \delta R_1^{\rm res}=-\delta R_1^{\rm
 res}=-\eta^{1/2}\zeta^{-1}a_d\cos(\phi_d+\beta).
\end{equation}
This expression shows that the correction to the activation energy
decreases as the system approaches the bifurcation point, i.e.,
$\eta=\mu-\mu_B^{(1)}$ decreases. However, the decrease of $\delta
R_1^{\rm res}$ is much slower than the decrease of $R^{(0)}$.
Therefore the relative correction to the activation energy sharply
increases as $\mu \to \mu_B^{(1)}$. It is important that the sign of
$\delta R_1^{\rm res}$, that shows which of the states 1 and 2 is
predominantly occupied, depends on $\phi_d-\beta$, that is, not only
on the relative phase of the additional field, but also on the
scaled relaxation parameter $\zeta$ that determines the value of
$\beta$. Therefore by varying $\zeta$ one can control which of the
states is predominantly occupied.

\subsection{High-frequency modulation}

We now consider the case where the additional field is detuned from
the subharmonic frequency, $|2\omega_d-\omega_F|\gg W_{12}^{(0)}$.
Since the motion near the bifurcation point is controlled by one
variable $Q'$, there is no phase shift between the components
$\chi_{1Q}(\nu_d)$ and $\chi_{1P}(\nu_d)$. The change of the
activation energy is $\delta R_1=\delta
R_2=-a_d\tilde\chi_{1c}(\nu_d)= -a_d|\tilde\cb_1(\nu_d)|$. The
problem as a whole coincides with that for a periodically modulated
overdamped equilibrium particle. The LS for an overdamped particle
in a quartic potential is already known \cite{Dykman2001}. In the
present case we have
\begin{eqnarray}
\label{eq:chi_near_mu_B}
 \tilde\chi_{1c}(\nu)&=&\pi^{-1/2}\zeta^{-1}\eta^{1/2}\nonumber\\
&&\times\left|\Gamma\Bigl((1-i\nu')/2\Bigr)\Gamma\bigl(1+i\nu'/2\bigr)\right|,
\end{eqnarray}
where $\nu'=\nu/|\mu_B^{(1)}|\eta\zeta$ and $\Gamma(z)$ is the Gamma
function. The function $\tilde\chi_{1c}(\nu)$ is proportional to a
smaller power of the distance to the bifurcation point $\eta$ than
$R^{(0)}$. This shows that the correction to the period-averaged
switching rate becomes relatively larger as the system approaches
the bifurcation point. As seen from Eq.~(\ref{eq:chi_near_mu_B}),
for small $\eta$ the LS $\tilde\chi_{1c}(\nu)$ has a peak at the
frequency detuning $\nu=0$ and monotonically decreases with
increasing $|\nu|$. The typical width of the peak of
$\tilde\chi_{1c}(\nu)$ is $\zeta|\mu_B^{(1)}|\eta$. It decreases as
$\mu$ approaches $\mu_B^{(1)}$.

\section{The weak damping limit}
\label{sec:weak_damping}

The LS can be analyzed also in the limit of weak damping, $\zeta\gg
1$. Here we consider damping that is weak in the rotating frame.
This means that not only is the oscillator decay slow compared to
its frequency $\omega_0$, but also compared to the much smaller
frequency $F/\omega_F$. If there were no damping and noise, the
motion of the oscillator in the rotating frame would be vibrations
with given quasienergy $g$, which are described by equation
$\dot\qb=\hat\epsilon\n g$. Damping causes the quasienergy to
decrease towards its value in one of the stable states whereas noise
leads to quasienergy diffusion away from these states. On the MPEP
$\qbon(\tau)$ the quasienergy increases from its value $g_n$ in the
stable state $n$ to its value at the saddle point $g_b$
\cite{Dykman1998} (the quasienergy $G$ in
Ref.~\onlinecite{Dykman1998} differs from $g^{(0)}(Q,P)$ in sign).
As a result, $\qbon(t)$ is a spiral.

\subsection{Symmetry lifting}
\label{subsec:static_response_weak_damping}

We will start the analysis with the case of the additional field at
exact subharmonic frequency, $\omega_d=\omega_F/2$.  In this case
$g(Q,P)$ is independent of time. The general expression for the
activation energy of switching in the limit of weak damping, which
is not limited to small $a_d$, has the form
\cite{Dykman1998,Marthaler2006,Dykman1979a}
\begin{eqnarray}
\label{eq:underdamped_static_general}
 &&R_n^{\rm
 res}=\zeta^{-1}\int\nolimits_{g_n}^{g_b} dg\,\frac{M_n(g)}{N_n(g)},\quad
  M_n(g)= \int\!\!\!\int\nolimits_{A_n(g)} dQdP, \nonumber\\
  &&  N_n(g)= \frac{1}{2}\int\!\!\!\int\nolimits_{A_n(g)} \n^2 g\, dQdP.
  \end{eqnarray}
Integration with respect to $Q,P$ in functions $M_n(g), N_n(g)$ for
given $g$ is done over the area $A_n(g)$ of phase plane $(Q,P)$,
which is limited by the phase trajectory $g(Q,P)=g$ that lies within
the basin of attraction to the stable state $\qb_n$.

For comparatively small $a_d$ the integrals in
Eq.~(\ref{eq:underdamped_static_general}) can be calculated by
perturbation theory. To first order in $a_d$ we obtain the following
expression for the correction $\delta R_n^{\rm res}\equiv \delta
R_n^{\rm res}(\phi_d)$ to the activation energy $R^{(0)}$ for
$\mu_B^{(1)} <\mu < \mu_B^{(2)}$,
\begin{eqnarray}
\label{eq:correction_underdamped}
 &&\delta R_1^{\rm res}(\phi_d)=-\delta R_2^{\rm res}(\phi_d) =
 {\cal X}_1\zeta^{-1}a_d\sin\phi_d,
 \end{eqnarray}
where
\begin{eqnarray}
\label{eq:X_1}
 {\cal X}_1&=&\int\nolimits_{g_{\min}^{(0)}}^0dg\frac{1}{N^{(0)}}
 \left[\delta M_1 -\frac{M^{(0)}}{N^{(0)}}\delta N_1\right] +\frac{(\mu
 +1)^{1/2}}{2+\mu},\nonumber\\
 \delta M_1&=&2\int\nolimits_{Q_{1\min}}^{Q_{1\max}} Q\,dQ/|\partial_Pg^{(0)}|,\\
 \delta N_1&=&\int\nolimits_{Q_{1\min}}^{Q_{1\max}}
 Q\,dQ\,\n^2g^{(0)}/|\partial_Pg^{(0)}|.\nonumber
 \end{eqnarray}
Here $M^{(0)}$ and $N^{(0)}$ are the values of $M(g)$ and $N(g)$ for
$a_d=0$, whereas $\delta M$ and $\delta N$ are the field-induced
corrections to $M$ and $N$ divided by $a_d\sin\phi_d$;
$g_{\min}^{(0)}=-(\mu+1)^2/4$. The limits $Q_{1\min}(g),
Q_{1\max}(g)$ of the integrals over $Q$ are given by equation
$g^{(0)}(Q,0)=g$, with $0< Q_{1\min} < Q_{1\max}$. The arguments of
the integrals over $Q$ in the expressions for $\delta M_1, \delta
N_1$ are calculated for $P$ given by equation $g^{(0)}(Q,P)=g$ with
$P>0$.

\begin{figure}[h]
\begin{center}
\includegraphics[width=3.0in]{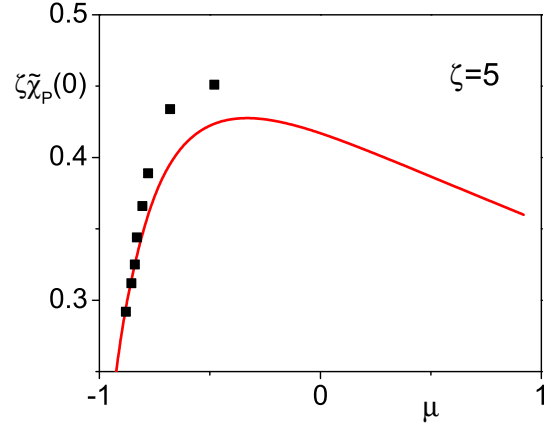}
\caption{Symmetry lifting in the weak-damping limit, $\zeta^{-1}\ll
1$. Solid line: the scaling factor ${\cal X}_1$, Eq.~(\ref{eq:X_1}),
in the correction to the activation energy $\delta R_1^{\rm
res}={\cal X}_1\zeta^{-1}a_d\sin\phi_d$. The dots show the scaled LS
$\zeta\tilde\chi_P(0)$. It gives the correction $\propto\sin\phi_d$
to $\delta R_1^{\rm res}$ and is obtained by calculating the MPEPs
for $\zeta = 5$.}
\label{fig:symm_lifting_underdamped}
\end{center}
\end{figure}

It follows from Eq.~(\ref{eq:correction_underdamped}) that, for
small damping, the dependence of the correction to the activation
energy on the phase of the additional field has a simple form
$\delta R_1^{\rm res}\propto \sin\phi_d$. Both $\delta R_1^{\rm
res}$ and $R^{(0)}$ are $\propto \zeta^{-1}$, they decrease with the
decreasing scaled friction coefficient $\zeta^{-1}$. The function
${\cal X}_1$ depends on one parameter, the relative frequency
detuning of the strong field $\mu$. This dependence is shown in
Fig.~\ref{fig:symm_lifting_underdamped}.

It is seen from Fig.~\ref{fig:symm_lifting_underdamped} that the
change of the activation energy is nonmonotonic as a function of
$\mu$. For $\mu$ close to the bifurcation value $\mu_B^{(1)}=-1$ we
have ${\cal X}_1\approx (\mu+1)^{1/2}$. This shows that the
weak-damping expression for $\delta R_1^{\rm res}$,
Eqs.~(\ref{eq:correction_underdamped}), (\ref{eq:X_1}), smoothly
goes over into expression (\ref{eq:deltaR_bif_static}) obtained near
the bifurcation point in the opposite limit of overdamped motion;
note that in Eq.~(\ref{eq:deltaR_bif_static}) $\beta\approx \pi/2$
for $\zeta^{-1}\ll 1$.

The analytical results are compared in
Fig.~\ref{fig:symm_lifting_underdamped} with the numerical results
obtained by directly solving the Hamiltonian equations of motion for
the MPEP and calculating the LS from Eqs.~(\ref{eq:deltaR_n}),
(\ref{eq:chi_omega}). Since in the limit of small damping $\delta
R_1^{\rm res}\propto \sin\phi_d$, one expects from
Eq.~(\ref{eq:staticR_n}) that $\tilde\chi_{1Q}(0)$ should be small
and ${\cal X}_1\approx\zeta\tilde\chi_{1P}(0)$. Close to the
bifurcation point $\mu_B^{(1)}$ the numerical results agree with the
asymptotic theory already for moderately small damping, $\zeta^{-1}
= 0.2$.

We found numerically that the component $\tilde\chi_{1Q}(0)$
increases with $\mu$. Close to the second bifurcation point
$\mu_B^{(2)}$ it becomes of the same order of magnitude as
$\tilde\chi_{1P}(0)$, for the chosen $\zeta$. We note that the MPEP
is a fast oscillating function of time, for weak damping. When
$\tilde\cb_1(0)$ is calculated, the oscillations largely compensate
each other. The numerical value of $\tilde\cb_1(0)$ is therefore
very sensitive to numerical errors in the MPEP. With increasing
$\mu$ the frequency of oscillations of the MPEP increases and so
does the sensitivity. Therefore we present numerical results only
for a limited range of $\mu$.

\subsection{Resonant peaks of the logarithmic susceptibility}
\label{subsec:resonant_LS_peaks}

Oscillations of the MPEP may be expected to lead to peaks of the LS
as function of frequency for the dimensionless frequency detuning
$|\nu|\gg \zeta^{-1}$. Since the period-averaged corrections to the
activation energy for the states 1 and 2, $\delta R_1$ and $\delta
R_2$, are equal, we will consider the MPEP for the state 1 and will
drop the subscript 1. The analysis of the LS peaks can be done by
extending to the MPEP of a parametrically modulated oscillator the
approach developed in Ref.~\cite{Dykman1979a}.

As a first step it is convenient to change variables from $Q,P$ to
quasienergy-angle variables $g,\psi$. This is accomplished by
seeking the optimal path of the unperturbed system, $a_d=0$, in the
form
\[\qb_{1\,\rm
opt}^{(0)}(\tau)=\bigl(x_{}(g_{\rm opt},\psi_{\rm opt}),y_{}(g_{\rm
opt},\psi_{\rm opt})\bigr).\]
Here $x_{},y_{}$ are the coordinate and momentum of vibrations with
given unperturbed quasienergy, $g^{(0)}(x_{},y_{})=g$; $\psi$ is
the vibration phase: $x_{}$ and $y_{}$ are $2\pi$-periodic in
$\psi$. We denote the vibration frequency by $\omega(g)$. The
equations for $x_{}, y_{}$ are of the form
\begin{equation}
\label{eq:x_psi_and_y_psi}
 \omega(g)\partial_{\psi}
x_{}=\partial_{y}g^{(0)},\qquad \omega(g)\partial_{\psi}
y_{}=-\partial_{x}g^{(0)}.
\end{equation}
Functions $x_{}(g,\psi)$ and $y_{}(g,\psi)$ can be expressed in
terms of the Jacobi elliptic functions \cite{Dykman1998}.

On the optimal path the quasienergy $g\equiv g_{\rm opt}(\tau)$ is a
function of time, with $\dot g_{\rm opt},\, |\dot\psi_{\rm
opt}-\omega(g_{\rm opt})|\propto\zeta^{-1}$. Using the explicit form
of equations of motion for the Lagrangian
(\ref{eq:switching_general}) one can show that, to first order in
$\zeta^{-1}$, on the optimal path
\begin{eqnarray}
\label{eq:opt_path_weak_damping}
 (dx/d\tau)_{\rm opt}&\approx& K_x^{(0)}(x,y) + \zeta^{-1}F(g)\partial_{\psi}y,\nonumber\\
 (dy/d\tau)_{\rm opt}&\approx&  K_y^{(0)}(x,y) - \zeta^{-1}F(g)\partial_{\psi}x.
 \end{eqnarray}
Here, the force ${\bf K}^{(0)}$, with components
$K_x^{(0)},K_y^{(0)}$, is defined by Eq.~(\ref{eq:K_force}) in which
$Q$ and $P$ are replaced by $x$ and $y$, respectively.

The function $F(g)$ in Eq.~(\ref{eq:opt_path_weak_damping}) remains
arbitrary, to first order in $\zeta^{-1}$. It can be found from the
analysis of the terms $\propto\zeta^{-2}$ in the Lagrange equation
for the optimal path \cite{Dykman1979a}, which is cumbersome. The
calculation can be simplified by noticing that $F(g)$ determines the
change of the quasienergy $\bar g$ averaged over vibration period
$2\pi/\omega(g)$. From Eq.~(\ref{eq:opt_path_weak_damping}), taking
into account the explicit form of ${\bf K}^{(0)}$ and expression
(\ref{eq:underdamped_static_general}), we have
\begin{eqnarray}
\label{eq:dot_g}
 && (d\bar g/d\tau)_{\rm opt} =-\zeta^{-1}\Bigl(\overline{x\partial_xg^{(0)}}
 +\overline
 {y\partial_yg^{(0)}}\nonumber\\
 &&+  F(\bar g)\omega^{-1}(\bar g)
 \left[\overline{(\partial_x g^{(0)})^2} +
 \overline{(\partial_y g^{(0)})^2}\right]\Bigr)\nonumber\\=
 &&-(\pi\zeta)^{-1}\left(\omega(\bar g)M^{(0)}(\bar g)+F(\bar g)N^{(0)}(\bar g)\right),
 \end{eqnarray}
where overline denotes period averaging.

Alternatively the evolution of quasienergy on the optimal path can
be found using the Langevin equation for $d\bar g/d\tau$, which can
be obtained by averaging Eq.~(\ref{eq:Langevin}) over the period
$2\pi/\omega(\bar g)$. The resulting equation describes drift and
diffusion of $\bar g$. The optimal path for $\bar g$ can be obtained
using the variational formulation similar to
Eq.~(\ref{eq:switching_general}). Since the corresponding
variational problem is one-dimensional, it is easy to find the
optimal path. As is often the case for one-dimensional systems
driven by white noise, the optimal path is the time-reversed path in
the absence of noise,
\[ \bigl(d\bar g/d\tau\bigr)_{\rm opt}=(\pi\zeta)^{-1}\omega(\bar
g)M^{(0)}(\bar g).\]
Comparing this expression with Eq.~(\ref{eq:dot_g}) we find
\begin{eqnarray}
\label{eq:F(g)}
 F(g)=-2\omega(g)M^{(0)}(g)/N^{(0)}(g).
 \end{eqnarray}
To lowest order in $\zeta^{-1}$ we can replace $\bar g$ with $g$ on
the optimal path.

Equations (\ref{eq:deltaR_n}), (\ref{eq:chi_omega}),
(\ref{eq:opt_path_weak_damping}), and (\ref{eq:F(g)}) describe the
LS in a simple form of the Fourier transform of the functions
$F(g)\partial_{\psi}x, F(g)\partial_{\psi}y$ on the optimal path.
The general expression is further simplified near the peaks of
$\tilde\cb_1(\nu)$. They occur where $\nu$ is close to the frequency
$\omega(g)$ or its overtones for certain values of $g$. It is
convenient to write $x_{}$ and $y_{}$ as Fourier series,
\[x_{}(g,\psi)=\sum\nolimits_nx_{}(m;g)\exp(im\psi)\]
and similarly for $y_{}$. Then calculating the LS is reduced to
taking the Fourier transform of the oscillating factors
$\exp[im\psi_{\rm opt}(\tau)]$ weighted with functions of $g_{\rm
opt}(\tau)$ that smoothly depend on time.

On the MPEP the leading term in the phase $\psi_{\rm opt}(\tau)$ is
$\psi_{\rm opt}(\tau)\approx
\int\nolimits^{\tau}d\tau'\omega\bigl(g_{\rm opt}(\tau')\bigr)$. The
function $\omega(g)$ monotonically decreases with increasing $g$.
Since $g_{\rm opt}(\tau) $ is monotonic as function of time,
$m\omega(g_{\rm opt})$ can be in resonance with $\nu$ on the
optimal path only at a certain time. This allows us to single out
resonant contributions $\tilde\cb_1(m;\nu)$ to $\tilde\cb_1(\nu)$
from the corresponding $m$th overtones of $x_{}(g,\psi),
y_{}(g,\psi)$. Near resonance integration over $\tau$ in
Eq.~(\ref{eq:chi_omega}) can be done by the steepest descent method,
for slowly varying $g(\tau)$. It gives
\begin{eqnarray}
 &&\tilde\cb_1(m;\nu)=C_m\pi\zeta^{-1/2}\left[2\nu M^{(0)}/
 |d\omega/dg|\,N^{(0)2}\right]_{g_{\nu m}}^{1/2}\nonumber\\
 &&\times\Bigl(y_{}(-m;g_{\nu m}),
 -x_{}(-m;g_{\nu m})\Bigr)\nonumber
 \end{eqnarray}
Here, $C_m$ is a phase factor, $|C_m|=1$. The subscript $g_{\nu m}$
indicates that the expression in the brackets should be calculated
for $g=g_{\nu m}$, with $g_{\nu m}$ given by the condition
$m\omega(g_{\nu m})=\nu$.

The change of the activation energy $\delta R_1$ is determined by
the LS $\tilde\chi_{1c}(\nu)$ defined in
Eq.~(\ref{eq:high-frequency}). From the explicit form of the matrix
elements $x(m;g), y(m;g)$ found in Ref.~\onlinecite{Marthaler2006}
it follows that $\arg[y(-m;g)x^*(-m;g)]=-\pi/2$ for $m>0$, in the
considered range of $g$ and $\mu$. Then from the expression for
$\tilde\cb_1(m;\nu)$ we obtain that the spectral peak of
$\tilde\chi_{1c}$ near an $m$th overtone, $\tilde\chi_{1c}(m;\nu)$,
is given by the expression
\begin{eqnarray}
\label{eq:chi_overtone_general}
 &&\tilde\chi_{1c}(m;\nu)=\pi\zeta^{-1/2}\left[2\nu M^{(0)}/
 |d\omega/dg|\,N^{(0)2}\right]_{g_{\nu m}}^{1/2}\nonumber\\
 &&\times\Bigl(|x_{}(-m;g_{\nu m})|+ |y_{}(-m;g_{\nu m})|\Bigr)\quad
 (m>0).
 \end{eqnarray}

The Fourier components $|x_{}(m;g)|, |y_{}(m;g)|$ rapidly decrease
with increasing $|m|$ (exponentially, for large $|m|$
\cite{Marthaler2006}). Therefore the major peak of
$\tilde\chi_{1c}(\nu)$ is the peak of $\tilde\chi_{1c}(1;\nu)$. From
Eq.~(\ref{eq:chi_overtone_general}), near the maximum it has the
form
\begin{equation}
\label{eq:chi1_asymptotic}
 \tilde\chi_{1c}(1;\nu)| \propto
(1-az)\theta(z), \qquad z=\omega(g_{\min})-\nu,
\end{equation}
where $a\sim 1$ is a numerical factor and $\theta(x)$ is the step
function [$\omega(g_{\min})=2(\mu+1)^{1/2}$]. In obtaining this
expression we used that, for small $g-g_{\min}$, vibrations with
given $g$ are nearly sinusoidal, with $x_{}(1;g),y_{}(1;g)\propto
(g-g_{\min})^{1/2}$ and $M^{(0)}, N^{(0)}\propto g-g_{\min}$.
The sharp asymmetry of the peak of $\tilde\chi_{1c}(1;\nu)|$ is due
to the fact that the eigenfrequencies $\omega(g)$ have a cutoff at
$\omega(g_{\min})$.

A similar calculation shows that the peak from the 2nd overtone is
smoother, with $\tilde\chi_{1c}(2;\nu)|\propto
[2\omega(g_{\min})-\nu]^{1/2}$ for small $2\omega(g_{\min})-\nu$.
The maximum of $\tilde\chi_{1c}(2;\nu)$ is shifted from
$2\omega(g_{\min})$ to lower frequencies and has a smaller height
than $\tilde\chi_{1c}(1;\nu)$. Higher-order peaks have still smaller
heights.

\begin{figure}[h]
\begin{center}
\includegraphics[width=3.0in]{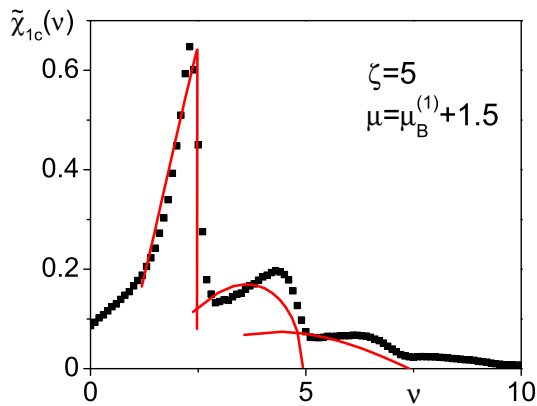}
\caption{The multiple-peak LS $\tilde\chi_{1c}(\nu)$ for an
underdamped system. Solid lines show the overtones
$\tilde\chi_{1c}(m;\nu)$, Eq.~(\ref{eq:chi_overtone_general}), with
$m=1,2,3$. Squares show the LS calculated by numerically finding the
MPEP. The data refer to $\zeta =5,\mu-\mu_B^{(1)}=1.5$. }
\label{fig:LS_peaks}
\end{center}
\end{figure}

The multi-peak structure of the LS for small damping is clearly seen
in Fig.~\ref{fig:LS_peaks}. The lines in this figure show asymptotic
expressions (\ref{eq:chi_overtone_general}) for the overtones
$\tilde\chi_{1c}(m;\nu)$. The squares are obtained by numerically
finding the optimal path $\qbo^{(0)}(t)$ and then calculating
$\tilde\chi_{1c}(\nu)$ from Eqs.~(\ref{eq:deltaR_n}),
(\ref{eq:chi_omega}), and (\ref{eq:high-frequency}). As expected
from Eq.~(\ref{eq:chi1_asymptotic}), the major peak of the
numerically calculated LS is at frequency $\omega(g_{\min})$. Other
peaks are located near overtones of $\omega(g_{\min})$, on the
low-frequency side. All peaks have characteristic strongly
asymmetric shapes.

Equation (\ref{eq:chi_overtone_general}) for the major peak, $m=1$,
is in a good agreement with the numerical calculations. The
agreement for the overtones is worse, because the peaks are much
broader and the contributions to the LS from different overtones
overlap. The full LS is not given just by a sum over $m$ of
$\tilde\chi_{1c}(m;\nu)$. It is necessary to take into account
interference of the contributions from vibrations with different
quasienergies $g_{\nu m}$ but close $m\omega(g_{\nu m})$ and, of
course, the effect of relaxation. We note that for $\nu\propto
2\omega_d-\omega_F <0$ the peaks of the LS have much smaller
amplitudes.

The above results demonstrate that, for an
underdamped oscillator, the rate of switching between period two
states can be resonantly increased by applying an extra field with
frequency $\omega_d\approx\omega_F/2$. The amplitude of the LS peaks
is $\propto \zeta^{-1/2}$, it is parametrically larger than the LS
at zero frequency, which determines the symmetry lifting and is
$\propto \zeta^{-1}$.

We note that, for extremely weak damping, a change of the switching
rate with the field amplitude $A_d$ may be associated with the
field-induced mixing of the attraction basins, as in equilibrium
systems \cite{Soskin2001b}. We do not discuss this mechanism here.

\section{Conclusions}
\label{sec:conclusions}

In this paper we considered an oscillator parametrically modulated
by a comparatively strong field at nearly twice its eigenfrequency
and additionally driven by a comparatively weak nearly resonant
field. Because of the parametric modulation the oscillator displays
period doubling. It has two vibrational states which differ only by
phase in the absence of the additional field. Even a comparatively
weak additional field can strongly affect the oscillator by changing
the rates of switching between the period two states. We have shown
that the rate change depends exponentially on the ratio of the
amplitude of this field $A_d$ to the characteristic fluctuation
intensity $D$. For small $D$ this change becomes large even where
the field only weakly perturbs the dynamics of the system.

If the frequency of the additional field $\omega_d$ coincides with
the frequency of the period two states $\omega_F/2$, the switching
rates $W_{12}$ and $W_{21}$ between the states become different from
each other. As a result, the stationary populations of the states
become different, too. This is the effect of symmetry lifting. It
depends exponentially strongly on the field amplitude.

For small frequency difference, $|\omega_d-\omega_F/2|\lesssim
W_{nm}$, the additional field leads to oscillations of the state
populations which, in turn, lead to fluctuation-induced three- and
multiple-wave mixing. For a larger detuning,
$|\omega_d-\omega_F/2|\gg W_{nm}$, the major effect of the
additional field is the increase of the switching rates $W_{nm}$
averaged over the beat period $4\pi/|\omega_F-2\omega_d|$.

Both for small and comparatively large $|\omega_F-2\omega_d|$ the
change of the switching rates is characterized by the LS
$\tilde\cb_n$. The latter gives the proportionality coefficient
between the field-induced change of the activation energy of
switching from an $n$th state $\delta R_n$ and the field amplitude
$A_d$. We have obtained an explicit general expression for the LS in
terms of the path that the system is most likely to follow in
switching, the MPEP. For $\omega_d=\omega_F/2$, the two components
of the vector $\tilde\cb_n$ give the coefficients in $\delta R_n$ at
$\cos\phi_d$ and $\sin\phi_d$, where $\phi_d$ is the phase of the
additional field relative to the phase of the strong field. For
$|\omega_d-\omega_F/2|\gg W_{nm}$ the quantity of interest is
$\tilde\chi_{1c}$ defined by Eq.~(\ref{eq:high-frequency}).

The major qualitative features of the LS are (i) scaling behavior
near the bifurcation point where the period two states merge, and
(ii) the occurrence of resonant peaks as a function of frequency for
weak damping. We have found that $\tilde\cb_n$ scales with the
distance $\eta$ to the bifurcation point as  $\tilde\cb_n \propto
\eta^{1/2}$. Thus the field-induced correction to the activation
energy of switching decreases as the system approaches the
bifurcation point. However, this decrease is significantly slower
than that of the major term in the activation energy, which is
$\propto\eta^2$.

Resonant peaks of the LS become pronounced where, in the rotating
frame, oscillator vibrations about the period two states are
underdamped. The vibration frequencies are much less than
$\omega_d$, and the condition that these vibrations are underdamped
is more restrictive than the requirement that the oscillator be
underdamped in the laboratory frame. The LS displays several peaks.
The major peak is shifted from $\omega_F/2$ by the frequency of
small-amplitude vibrations about the period two states. Other peaks
are shifted approximately by the overtones of this frequency and
have smaller amplitudes. All LS peaks have characteristic strongly
asymmetric shapes.

The effects discussed in this paper are not limited to a parametric
oscillator, they can be observed in other systems with period two
states. The results on modulation of switching rates by an
additional field can be extended also to other systems with
coexisting vibrational states, including resonantly driven nonlinear
nano- and micromechanical oscillators and Josephson junctions.
Fluctuational interstate transitions in these systems were recently
studied experimentally
\cite{Siddiqi2004,Aldridge2005,Stambaugh2006}. Although the theory
of the LS for modulated oscillators was not developed until this
paper, a resonant increase of the switching rate by an additional
field was expected from qualitative arguments based on the analogy
with static systems \cite{Dykman_neverpublished}. The preliminary
experimental data indicate that the effect occurs in Josephson
junctions when the frequencies of the both strong and weak fields
are close to the plasma frequency \cite{Siddiqi_private}.

In conclusion, we have studied the effect of an additional field on
a fluctuating parametrically modulated oscillator. We predict strong
change of the populations of the period two states by a
comparatively weak field. We also predict that the logarithm of the
rate of interstate switching is linear in the field amplitude, and
the proportionality coefficient may display resonant peaks as a
function of the field frequency.

MID is grateful to  M. Devoret, M. S. Heo, W. Jhe, and I. Siddiqi
for stimulating discussions. This work was supported by the NSF
through grant No. ITR-0085922 and by the ARO through grant No.
W911NF-06-1-0324.


\end{document}